# Cavity-enhanced superconductivity in the two-dimensional limit of $NbSe_2$

Hanxiang Zhang[1,#], Zexin Feng[2,#], I-Te Lu[3,4], Zhiwei Li[2], Songhao Guo[2], Qiuyu Shang[2], Thomas Tan[5,6], Xiaodan Lyu[2], Xiangming Shen[2], Dening Luan[6,7], Mingcheng Panmai[2], Kenji Watanabe[8], Takashi Taniguchi[9], Ranjan Singh[10], Angel Rubio[3,11] & Weibo Gao[1,2,6,7]*

1. Centre for Quantum Technologies, Singapore.
2. School of Electrical and Electronic Engineering, Nanyang Technological University, Singapore 639798, Singapore.
3. Max Planck Institute for the Structure and Dynamics of Matter and Center for Free-Electron Laser Science, Luruper Chaussee 149, Hamburg 22761, Germany.
4. National Yang Ming Chiao Tung University, Hsinchu 300093, Taiwan.
5. Centre for Disruptive Photonic Technologies, The Photonics Institute, Nanyang Technological University, Singapore 637371, Singapore.
6. Division of Physics and Applied Physics, School of Physical and Mathematical Sciences, Nanyang Technological University, Singapore 637371, Singapore.
7. Quantum Science and Engineering Center (QSec), Nanyang Technological University, Singapore 639798, Singapore.
8. Research Center for Electronic and Optical Materials, National Institute for Materials Science, Tsukuba, Japan.
9. Research Center for Materials Nanoarchitectonics, National Institute for Materials Science, Tsukuba, Japan.
10. Department of Electrical Engineering, University of Notre Dame, Notre Dame, IN, USA.
11. Initiative for Computational Catalysis (IIC) and Center for Computational Quantum Physics (CCQ), The Flatiron Institute, 162 Fifth Avenue, New York, NY 10010, USA.

#These authors contributed equally to this work.

* Corresponding author. Email: wbgao@ntu.edu.sg

**Abstract:** Vacuum electromagnetic fluctuations have emerged as a means of controlling collective quantum phases without external driving. Cavity-induced modification of superconductivity has been widely predicted. What sets the size of the effect, and which microscopic channel carries it, remain open. Here we couple few-layer $NbSe_2$ to a terahertz complementary split-ring resonator (CSRR) and show that the enhancement grows sharply on approaching the two-dimensional limit. In bilayer $NbSe_2$ the superconducting transition temperature rises by ~10%, from 3.02 K to 3.41 K, on a cavity resonant at 0.92 THz — roughly four times the shift measured in a ten-layer device at the same resonance. Within a single device the shift maps onto the simulated cavity field profile, falling from 0.39 K at the field maximum to zero outside the resonator, with the lower critical field following the same spatial ordering; because all regions are measured on one continuous flake in a single cooldown, sample-to-sample variation is excluded by construction. The frequency dependence is non-monotonic, with suppression below resonance and maximal enhancement near 0.96 THz. Quantum electrodynamical density functional theory calculations show that cavity coupling redistributes spectral weight in the Eliashberg function, weakening the total electron-phonon coupling while hardening the logarithmic average phonon frequency; competition between the two reproduces a sign change in $\Delta T_c$. These results identify dimensionality, local field amplitude and detuning as the control parameters of cavity-enhanced superconductivity, and point to electron-phonon reweighting as its microscopic origin.

## Introduction

In quantum electrodynamics, vacuum is not void but permeated by fluctuations of the electromagnetic field, evinced by Lamb shift[1], the Casimir effect[2] and spontaneous emission. Although contributions from these vacuum fluctuations were previously neglected in condensed matter physics due to their relatively small amplitude, recent studies have demonstrated that as ultra-strong coupling between vacuum fields and matter is achieved, control of material ground state is possible[3-6]. Specifically, theoretical studies have predicted that cavity electromagnetic fields can alter collective quantum phases, leading to phenomena such as engineered ferroelectricity[7-9], modified magnetic and topological phases[10,11], and tunable superconductivity[12-17]. Correspondingly, a growing number of experiments have revealed cavity-induced control of quantum Hall states[18-22], photon-mediated attractive interactions[23], charge-density-wave systems[24], magnetic nanoparticles[25], and superconductors[26-28]. These advances establish electromagnetic-environment engineering as a powerful approach for manipulating quantum materials beyond conventional methods.

Among various quantum phases, superconductivity—an emergent ground state characterized by macroscopic quantum coherence—has attracted considerable attention in this context. Theoretical studies have proposed several mechanisms for enhancing superconductivity via light–matter interaction, including modified electron-phonon coupling and phonon frequency via cavity-induced change in electron density[12,29,30], direct electron-photon coupling[13,31] and band engineering[32]. Experimental efforts have probed cavity-induced effects on superconducting properties, including measurements of the superfluid density in κ-ET[26] and of the superconducting gap in NbN[27], and evidence of an enhanced superconducting response in YBCO[28]. While the present work was in preparation, transport evidence for a cavity-enhanced transition temperature was reported independently in 12-layer $NbSe_2$ embedded in a metallic split-ring resonator[33]. That result establishes the effect in the thick-flake regime and, together with the measurements reported here, places it on firm experimental footing. What remains open is how the enhancement scales as the superconductor approaches the two-dimensional limit, whether it can be resolved spatially against the cavity field profile within a single device, and which microscopic channel is responsible.

In this work, we report cavity-enhanced superconductivity in bilayer and ten-layer $NbSe_2$ coupled to a complementary split-ring resonator (CSRR), and organize the results around three observations. First, the effect depends strongly on dimensionality: for a CSRR resonant at $\omega_c$ = 0.92 THz, the superconducting transition temperature ($T_c$) of the bilayer rises from 3.02 K to 3.41 K, an increase of ~10%, whereas an identically prepared ten-layer device on the same resonance shifts by only 0.10 K. Second, the enhancement can be resolved spatially within a single device: using multiple voltage probes across one CSRR we find the largest shift where the simulated field is maximal ($\Delta T_c$ = 0.39 K), a reduced shift toward the edge of the gap (0.27 K), and no measurable shift outside the resonator, with the lower critical field following the same ordering. Because these regions lie on one continuous flake measured in a single temperature sweep, extrinsic sample-to-sample variation is excluded by construction. Third, the response is resonant: superconductivity is suppressed at 0.69 THz, maximally enhanced near 0.96 THz, and unaffected at 2.00 THz. Quantum electrodynamical density functional theory calculations trace this non-monotonic behaviour to a redistribution of the Eliashberg spectral function among existing phonon branches. Together these results move the question from whether vacuum fields can enhance superconductivity to what controls the magnitude and sign of the enhancement.

## Results and Discussion

### Cavity enhanced superconductivity

Figure 1a presents the resistance versus temperature ($R$-$T$) curve of bilayer $NbSe_2$ which shows a sharp transition from a normal resistive state to a superconducting state as temperature decreases. The inset depicts a zoomed-in view of the transition region, where the resistance drops steeply to zero. Here, we define the superconducting transition temperature ($T_{c,0}$) as where the resistance decreases by two orders of magnitude. In this bilayer $NbSe_2$, the $T_{c,0}$ is 3.02 K in the absence of CSRR, consistent with previous studies[34,35].

Based on the superconducting transition of pristine bilayer $NbSe_2$, we investigated the influence of the CSRR on its superconductivity. As illustrated in Figure 1b and Extended Figure 1a, the device consists of a bilayer $NbSe_2$ flake with thickness of 1.98 nm (Extended Fig. 1b, c), placed on a $SiO_2$/Si substrate, contacted by Cr/Au electrodes for electrical measurements. Layer number of the flake is subsequently verified by employing second-harmonic generation (SHG), see Extended Fig. 1d, e for more details. An insulating hexagonal boron nitride (hBN) capping layer of ~30 nm situates above the $NbSe_2$ sheet to protect it from air degradation and separate it from the gold CSRR on top. In this scenario, the CSRR acts as the vacuum cavity that exerts vacuum field fluctuations on $NbSe_2$ without direct electrical contact.

The CSRR is characterized by terahertz time-domain spectroscopy (THz-TDS). As shown in Figure 1c, the transmission spectrum displays a dip in the middle, yielding a resonant frequency at $\omega_c$ = 0.92 THz.

Transport measurements reveal an enhancement of the superconducting transition temperature in the presence of the CSRR. A clear shift is observed between the *R*-*T* curves with and without the CSRR, demonstrating an enhancement of $T_{c,0}$ by 0.39 K (Fig. 1d). In few-layer $NbSe_2$, the superconducting transition temperature depends sensitively on layer thickness[35]. Therefore, variations in thickness could lead to apparent differences in $T_{c,0}$. To exclude the sample thickness induced variation of $T_{c,0}$, we fabricated a control sample with conspicuous layer difference (see Extended Fig. 2a). Using the same electrode geometry as in Figure 1b, all electrode pairs exhibit a ladder-like transition, reflecting multiple superconducting regions with different $T_c$ values (see Extended Fig. 2b). In contrast, all neighboring electrode pairs in our devices show single-step transitions without any ladder-like features, indicating a high thickness uniformity (Extended Fig. 2c). If thickness variations were present, at least one electrode pair would cross a layer boundary and exhibit a multi-step transition. These results rule out sample inhomogeneity as the origin of the observed $T_c$ enhancement.

It should be mentioned that in $NbSe_2$, the charge density wave (CDW) phase is generally considered a competing order with superconductivity and tends to suppress it[36,37]. An alternative explanation of this superconductivity enhancement could be that vacuum fluctuations constrain the CDW phase. To rule out possible contribution from CDW, Raman spectroscopy is performed on regions with and without CSRR in the normal state (~4 K). The results show that the variations of the soft mode, $A_{1g}$ and $E_{2g}$ peaks in $NbSe_2$ inside and outside the cavity are negligible (see Extended Fig. 3). The soft mode is closely associated with the CDW order parameter, whereas the $A_{1g}$ and $E_{2g}$ modes originate from intrinsic lattice vibrations and primarily reflect the structural properties of the crystal[38]. These observations suggest that the CDW contribution is unlikely to play a dominant role.

Another possible trivial contribution comes from the gold CSRR. Gold could potentially screen the Coulomb interaction between electrons in $NbSe_2$ beneath, and affect superconductivity. We picked an area in our sample and evaporated a gold square onto it. We measured *R*-*T* curves in regions covered by the gold square instead of gold CSRR, and the results are identical as the case without CSRR (Extended Fig. 4a). Also, thermalization is a factor that cannot be neglected, since the gold cavity could introduce delayed thermalization. We swept temperature downwards and upwards repetitively and found that the two curves are highly identical (Extended Fig. 4b). These two control experiments rule out alternative explanation that the enhanced superconductivity comes from gold itself, but from the cavity structure.

Taken together, these results allow us to attribute the observed enhancement of superconductivity to the resonant electromagnetic environment introduced by the CSRR under appropriate conditions.

**Dependence of $T_{c,0}$ enhancement on resonant field strength**

Figure 2a presents the simulated electric field distribution of the CSRR used in Figure 1b, showing a pronounced field enhancement at the center that gradually decreases toward the periphery. The employed 0.92 THz CSRR has a length of 30 μm and a width of 8 μm. The electric field inside the cavity gap is uniformly distributed with an amplitude of approximately 0.8 V/m, while it decays rapidly on both ends of the gap, vanishing at a distance of about 4 μm from the edge. This spatial variation enables us to explore the effect of field strength on superconductivity via in situ measurements across different regions within the same device.

To exploit this spatially varying field profile, we performed electrical measurements using multiple voltage probes, namely P1, P2, and P3 (Fig. 2b) in the same device, enabling spatially resolved characterization of superconducting transition. The *R*-*T* curves measured at different probe locations exhibit slightly different transition temperatures, evidencing a position-dependent modification of superconductivity (Fig. 2c). The largest enhancement is observed at P1 ($\Delta T_c$ = 0.39 K), where the field amplitude is maximal, while a reduced enhancement is found at P2 ($\Delta T_c$ = 0.27 K), consistent with the weaker vacuum field. In contrast, no measurable enhancement is detected at P3, located outside the CSRR region.

We further examine the lower critical field ($H_{c1}$) as a function of temperature, which is shown in Figure 2d. In all regions, $H_{c1}$ decreases approximately linearly with increasing temperature, as expected for type-II superconductors. Notably, the absolute values of $H_{c1}$ follow the same spatial trend as $T_{c,0}$, with larger values observed in regions of stronger field enhancement (Fig. 2d). This result indicates that the superconductivity of $NbSe_2$ can be enhanced inside the cavity.

These results establish a clear correlation between the local electromagnetic field strength and the superconducting properties, indicating that the enhancement of superconductivity is governed by the magnitude of the resonant vacuum electromagnetic field. Moreover, since all measurements are performed within the same device, extrinsic sample-to-sample variations can be excluded, further supporting an intrinsic field-induced origin.

**Dependence of $T_{c,0}$ enhancement on frequency**

To elucidate the role of the cavity resonance, we further performed measurements under different resonant cavity frequencies. We designed 5 CSRRs of different resonant frequencies (see schematic in Fig. 3a). Since we can only discern terahertz signal from substrate noise under 1.5 THz, we measured transmission spectrum of three CSRRs with resonant frequency lower than 1.5 THz, and extrapolated that of other two using the fitted data (Extended Fig. 5). Limited by sample sizes to incorporate larger area of the lower frequency cavity, we utilized the ten-layer sample instead for experiments below. Figure 3b-d present results with vacuum fields at 0.69, 0.92 and 2.00 THz coupled to ten-layer $NbSe_2$, respectively. At 0.69 THz, the superconducting transition shifts slightly to lower temperatures with $\Delta T_{c,0}$ of -0.12±0.01 K, indicating that low-frequency excitation disrupts superconducting coherence, possibly by reducing the superconducting gap of the Cooper pairs. At 0.92 THz, the ten-layer $NbSe_2$ device still exhibits an enhancement of superconductivity, although the increase in $T_{c,0}$ is limited to 0.10±0.01 K, significantly smaller than that observed in the bilayer device. Similar tiny enhancement in $T_{c,0}$ is seen in other different two measurements (see Extended Fig. 6a-c), indicating the reproducibility of our experiments. By contrast, at 2.00 THz, the superconducting transition curves with and without the CSRR nearly overlap, suggesting that the cavity exerts little influence on superconductivity at this frequency.

Notably, the dependence of $\Delta T_{c,0}$ on frequency is non-monotonic. As summarized in Figure 3e, $\Delta T_{c,0}$ initially increases at lower frequencies, reaches a maximum enhancement near ~0.96 THz, and then decreases at higher frequencies. This behavior indicates that the superconducting response is highly sensitive to the resonant electromagnetic environment, with an optimal frequency window where the coupling enhances superconductivity, while lower and higher-frequency excitations lead to suppression and no response, respectively.

It is instructive to compare these observations with the concurrent, independent report of cavity-enhanced superconductivity in 12-layer $NbSe_2$ coupled to a metallic split-ring resonator[33]. The two experiments differ in resonator topology—a Babinet-complementary aperture here versus a metallic split ring there—and in sample thickness, yet they agree on the essential phenomenology: a resonant enhancement of $T_c$, suppression on the low-frequency side of resonance, and an optimum near 0.95 THz there against 0.96 THz here. Agreement of the optimal frequency across two different field geometries, sample batches and laboratories is a stringent test of the cavity-electrodynamic origin of the effect, and argues against device-specific or fabrication-related artefacts in either experiment. The two datasets are also quantitatively consistent where they overlap in thickness: our ten-layer device gives $\Delta T_{c,0}$ = 0.10 ± 0.01 K, comparable to the 0.15 K reported at 12 layers. The bilayer measurement reported here lies a factor of four higher in relative terms and identifies dimensionality as a strong tuning parameter, consistent with the expectation that enhanced superconducting fluctuations at reduced thickness strengthen the low-energy electromagnetic response and therefore the coupling to the cavity mode.

Finally, we consider a possible interpretation of these results among other potential explanations[28,39], while noting that the exact microscopic origin of the cavity-induced changes in $T_c$ remains unclear. Recent first-principles work on cavity-modified superconductivity in $MgB_2$ has suggested, within quantum electrodynamical density functional theory (QEDFT), that coupling to cavity photon modes can alter the electronic ground state and thereby modify the phonon spectrum[30]. Guided by this idea, we extended our QEDFT calculations for monolayer 2H-$NbSe_2$ in the non-CDW phase over a broad range of effective electron-photon coupling strengths, parameterized by the dimensionless ratio $\frac{\lambda_\alpha}{\omega_\alpha}$ of the effective light-matter coupling $\lambda_\alpha$ to the effective photon frequency $\omega_\alpha$, and evaluated the resulting Allen-Dynes $T_c^{AD}$ for each (Fig. 4a). The calculated shift $\Delta T_c^{AD}$ is not monotonic; it is positive over most of the range explored but reverses sign in an intermediate window ( $\frac{\lambda_\alpha}{\omega_\alpha} \approx 0.03 - 0.08$ ). Decomposing $\Delta T_c^{AD}$ into the underlying Eliashberg quantities reveals the origin of this behavior. Increasing photon coupling decreases the total electron-phonon strength $\lambda$ while simultaneously increasing the logarithmic average phonon frequency $\omega_{log}$ (Fig. 4b, c). These two quantities enter the Allen-Dynes formula with opposite effects on $T_c$, and their contributions are comparable in magnitude over much of the range explored; the net shift therefore reflects a close competition between them, providing a physically transparent origin for the sign reversal.

To identify the microscopic origin of this competition, we compared the phonon density of states (DOS) and the Eliashberg spectral function $\alpha^2F(\omega)$ for two representative coupling strengths, $\frac{\lambda_\alpha}{\omega_\alpha} = 0.005$ and 0.03, chosen to correspond to enhanced and suppressed $T_c$, respectively (Fig. 4d, e). The phonon DOS is essentially unchanged from the uncoupled case across the full frequency range, indicating that photon coupling does not appreciably renormalize the interatomic force constants. In contrast, $\alpha^2F(\omega)$ is markedly redistributed in frequency; spectral weight shifts toward the low-frequency phonon branches (below ~2 THz) in the enhanced case, and toward higher-frequency optical modes (~5-6 THz) in the suppressed case. Sine $\lambda$ weights $\alpha^2F(\omega)$ by $1/\omega$ and is therefore dominated by low-frequency modes, whereas $\omega_{log}$ is a logarithmic average that rises as weight moves to high frequencies, this redistribution accounts directly for the concurrent decrease in $\lambda$ and increase in $\omega_{log}$ identified above. Cavity coupling thus acts primarily on the electron-phonon matrix elements themselves, redistributing coupling strength among the existing phonon modes, rather than by renormalizing the phonon spectrum.

We emphasize that these calculations, performed for monolayer $NbSe_2$ in the non-CDW phase, are not intended as a quantitative model of the bilayer and ten-layer devices measured here, and that the coupling-strength parameter does not map directly onto the cavity frequency varied in Fig. 3. They nonetheless demonstrate that a competition between cavity-induced weakening of the electron-phonon coupling and hardening of the characteristic phonon energy scale offers a physically consistent, qualitative rationale for how vacuum photon coupling can drive $T_c$ in either direction, in line with the non-monotonic frequency dependence of $\Delta T_{c,0}$ observed experimentally (Fig. 3e). We further note that the screened Coulomb pseudopotential, $\mu^*$, was held fixed at 0.15 across all coupling strengths; in principle, cavity coupling could also renormalize the electronic screening that determines $\mu^*$, which is not captured by the present calculations. A quantitative account that includes both the correct layer number and the CDW phase, together with a self-consistent treatment of $\mu^*$ and a direct mapping between cavity frequency and mode-resolved coupling strength, remains an important goal for future work. We further note that the electron-phonon route examined here is complementary to the Ginzburg-Landau analysis advanced for the thick-flake experiment[33], in which the paramagnetic exchange contribution is bounded by the diamagnetic term for a passive minimal cavity[39], so that enhancement requires additional collective degrees of freedom that are treated phenomenologically. Cavity-induced reweighting of the electron-phonon matrix elements provides one concrete microscopic realization of such a channel, and the two descriptions address different sectors of the same problem rather than competing accounts of it.

## Conclusion

In summary, we have shown that coupling few-layer $NbSe_2$ to a complementary split-ring resonator modulates superconductivity through a resonant THz electromagnetic environment, and that the magnitude and sign of the modulation are set by three controllable parameters. Reducing the thickness to the bilayer limit raises the enhancement of $T_{c,0}$ to ~10% at 0.92 THz, roughly four times that of a ten-layer device; moving the voltage probe across the resonator maps the shift onto the local vacuum field amplitude within a single flake; and detuning the resonance converts enhancement into suppression. Taken together with the concurrent observation in thicker flakes[33], these results place cavity-enhanced superconductivity on firm experimental footing and establish electromagnetic-environment engineering as a quantitatively controllable route for tuning superconductivity, with the two-dimensional limit as the regime in which it is most effective. This technique is readily generalizable to a broad class of correlated electron systems that host superconductivity and other emergent phases, such as twisted bilayer graphene[40], ABC-stacked graphene[41], and twisted $MoTe_2$[42]. By combining metamaterial resonators with quantum materials, a versatile platform is established for exploring cavity-engineered quantum states and nonequilibrium light–matter interactions, creating new avenues for tailoring collective quantum phenomena and realizing novel quantum technologies.

## Methods

### Device fabrication

Prepatterned electrodes were fabricated on $Si/SiO_2$ substrates with a 285 nm $SiO_2$ layer. The electrodes were defined by electron-beam lithography (EBL), followed by thermal evaporation of 2 nm Cr and 30 nm Au. $NbSe_2$ flakes were exfoliated onto polydimethylsiloxane (PDMS) from a commercial single crystal purchased from HQ Graphene. The thickness of representative flakes was first determined by atomic force microscopy (AFM), and the corresponding optical contrast was also obtained, allowing subsequent identification of flakes with desired thickness directly from optical images. The hBN flakes, typically ~30 nm thick, were mechanically exfoliated onto Si/$SiO_2$ substrates and their thickness was identified by optical contrast. The desired hBN flakes were then picked up using the polycarbonate (PC)/PDMS stamp. The hBN layer serves both as an encapsulation layer to protect $NbSe_2$ from oxidation and as an insulating spacer preventing direct electrical contact between $NbSe_2$ and the CSRR. Two device fabrication routes were employed. In the first route, $NbSe_2$ flakes exfoliated on PDMS were transferred onto Si/$SiO_2$ substrates and subsequently patterned into uniform rectangular strips using an AFM tip. The patterned flakes were then picked up by hBN on the PC/PDMS stamps and transferred onto the prepatterned electrodes. In the second route, $NbSe_2$ flakes were first transferred directly onto the electrode substrates and then shaped into rectangular strips using an AFM tip, followed by encapsulation with hBN on the PC/PDMS stamp. All exfoliation, transfer, and encapsulation processes related to $NbSe_2$ were carried out in an argon-filled glovebox with $H_2O$ and $O_2$ levels below 0.1 ppm. After hBN encapsulation, the device was taken out of the glovebox, and the CSRR was defined by EBL. Finally, 2 nm Cr and 30 nm Au were deposited by thermal evaporation to form the CSRR.

**THz-TDS measurements**

An 800 nm laser pulse is generated from a Ti:Sapphire amplifier with a repetition rate of 1 kHz, a pulse-width of ~45 fs and power of ~4 W, is used to generate THz pulses from ZnTe crystal. Through nonlinear optical rectification and electro-optic sampling, THz time domain spectroscopy is conducted.

The time domain signal through the sample and reference is measured, $E_{Sam}(t)$ and $E_{Ref}(t)$. A Fast Fourier Transform (FFT) is done to the time domain signal to obtain the frequency dependent THz signal, $E_{Sam}(\omega)$ and $E_{Ref}(\omega)$. A normalization against a reference background is done to obtain the transmission spectrum, $T(\omega)$.

$$T(\omega) = \left|\frac{E_{Sam}(\omega)}{E_{Ref}(\omega)}\right|$$

**Electrical transport measurements**

Electrical transport measurements were performed in a Quantum Design Physical Property Measurement System (PPMS). Temperature-dependent superconducting transition was conducted using a low-frequency AC method. An excitation current of 1 μA at 17.777 Hz was supplied by a Stanford Research Systems SR830 lock-in amplifier and the voltage signal was detected using Zurich Instruments MFLI lock-in amplifiers.

The temperature sweep rate was maintained at 0.2 K/min in the vicinity of the superconducting transition. For each device, the superconducting transitions of regions with and without the CSRR were measured during the same temperature sweep, ensuring identical experimental conditions.

**Raman measurements**

Raman spectra were measured on a home-built low-wavenumber Raman system using a 532 nm laser (08-01, Cobolt). The sample was mounted in a cryostat (AttoDry 1000) at 4 K. The laser power was kept at 0.5 mW. The Raman signal was collected by a spectrograph (HRS-500, Princeton Instruments) coupled to a nitrogen-cooled detector (PyLon100BR_eXcelon, Princeton Instruments).

**First-principles calculations**

Density functional theory calculations for monolayer 2H-$NbSe_2$ with the non-CDW phase were performed using Quantum ESPRESSO[43,44] with optimized norm-conserving Vanderbilt pseudopotentials from the PseudoDojo library[45] and the Perdew-Burke-Ernzerhof exchange-correlation functional[46]. The kinetic-energy cutoff for the wavefunctions was set to 80 Ry. Brillouin-zone integrations were carried out using Marzari-Vanderbilt smearing[47] with a smearing width of 0.02 Ry to avoid imaginary phonons and a Γ-centered Monkhorst-Pack $24 \times 24 \times 1$ **k**-point mesh. The primitive cell was relaxed until the residual forces were below $10^{-5}$ Ry/Bohr, giving optimized lattice constants of $a = 3.47137$ Å with a vacuum space of 20.0 Å. Spin-orbit coupling and van der Waals corrections were not included. Phonon properties were calculated using density functional perturbation theory as implemented in the PHonon package of Quantum ESPRESSO[43,44]. The dynamical matrices were computed on a Γ-centered $12 \times 12 \times 1$ **q**-point mesh and Fourier transformed to obtain the interatomic force constants, which were then used to interpolate the phonon dispersion. To account for the coupling between the electronic system and quantized cavity photon modes, we performed quantum electrodynamical density functional theory (QEDFT) calculations[30,48]. In this framework, the light-matter interaction is characterized by the user-defined collective coupling strength $\lambda_\alpha$ and bare photon frequency $\omega_\alpha$. In the present work, one photon mode with the in-plane polarization aligned with the Nb-Se bond direction was included. The ground-state and phonon calculations for the photon-coupled system were performed using the same computational settings as in the uncoupled case. To compute total electron-phonon coupling (and then Allen-Dynes $T_c$), we used the EPW package to interpolate electron-phonon matrix elements with the fine k-point grid of $120 \times 120 \times 1$ and the fine q-point grid of $60 \times 60 \times 1$. The cutoff for the acoustic phonon frequencies was set to 16 $cm^{-1}$, and the broadening value for the electron (phonon) system was set to 25 (0.05) meV.

**Data availability:** Source data are provided with this paper.

## Acknowledgements

We sincerely thank Cristiano Ciuti and Itai Keren for fruitful discussions. This work is supported by ASTAR (M24M8b0004), Singapore National Research foundation (NRF-CRP30-2023-0003, NRF-CRP31-0001, NRF2023-ITC004-001 and NRF-MSG-2023-0002) and Singapore Ministry of Education Tier 2 Grant (MOE-T2EP50222-0018).

## Author contributions

H.Z. and W.G. conceived the idea. H.Z. designed the CSRR with assistance from D.L. and M. P. H.Z. and Z.F. fabricated the devices with assistance from Z.L. and X.S. H.Z. and Z.F. performed the transport measurements and data analysis. S.G. and Q.S. carried out the Raman measurements and analysis. T.T. and R.S. performed the THz-TDS measurements. X.L. performed the SHG measurements. A.R. and I.L. provided theoretical input and interpretation. K.W. and T.T. provided high-quality hBN crystals. H.Z., Z.F. and I.L. prepared the paper. All authors discussed the results and commented on the manuscript. The project was led by W.G.

## Competing interests

The authors declare no competing financial interests.

## Figure 1

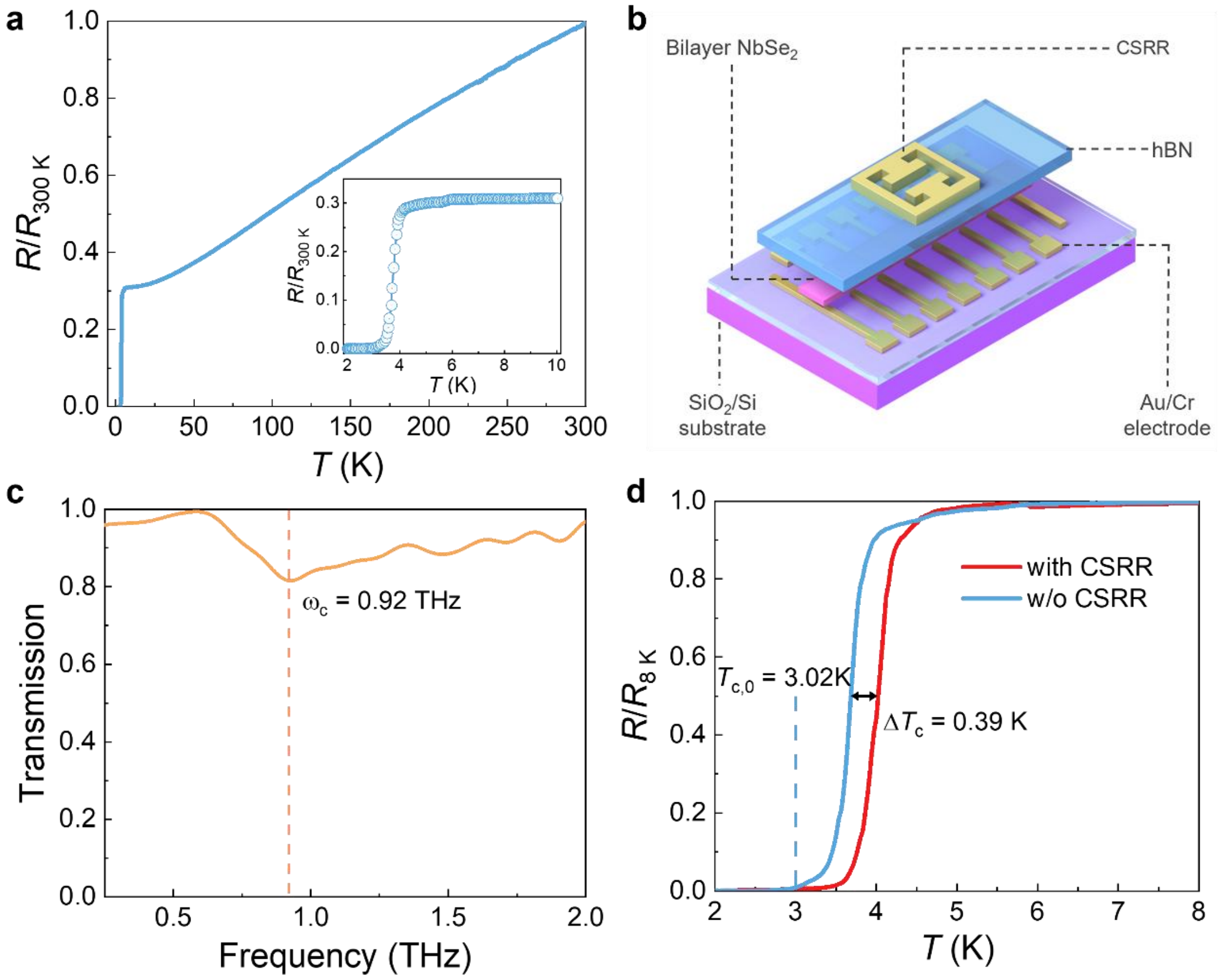

**Figure 1 | Enhancement of superconductivity in $NbSe_2$ coupled to a complementary split-ring resonator (CSRR). a**, Normalized resistance-temperature (*R*-*T*) curve of a pristine bilayer $NbSe_2$ measured from 300 to 2 K. Right inset is the zoom-in version of data between 10 to 2 K. The superconducting transition temperature, defined by the zero-resistance temperature ($T_{c,0}$), is 3.02 K. **b**, Schematic of a CSRR integrated with a $NbSe_2$ device in a four-probe measurement configuration. A ~30-nm hBN is inserted to protect $NbSe_2$ and prevent direct electrical contact with CSRR. **c**, Measured THz-TDS transmission spectrum on CSRR. The transmission spectrum displays a dip near 1 terahertz, yielding $\omega_c$ = 0.92 THz. **d**, *R*-*T* curves measured in the regions with and without CSRR. The corresponding $T_{c,0}$ values are 3.41 K and 3.02 K, respectively.

## Figure 2

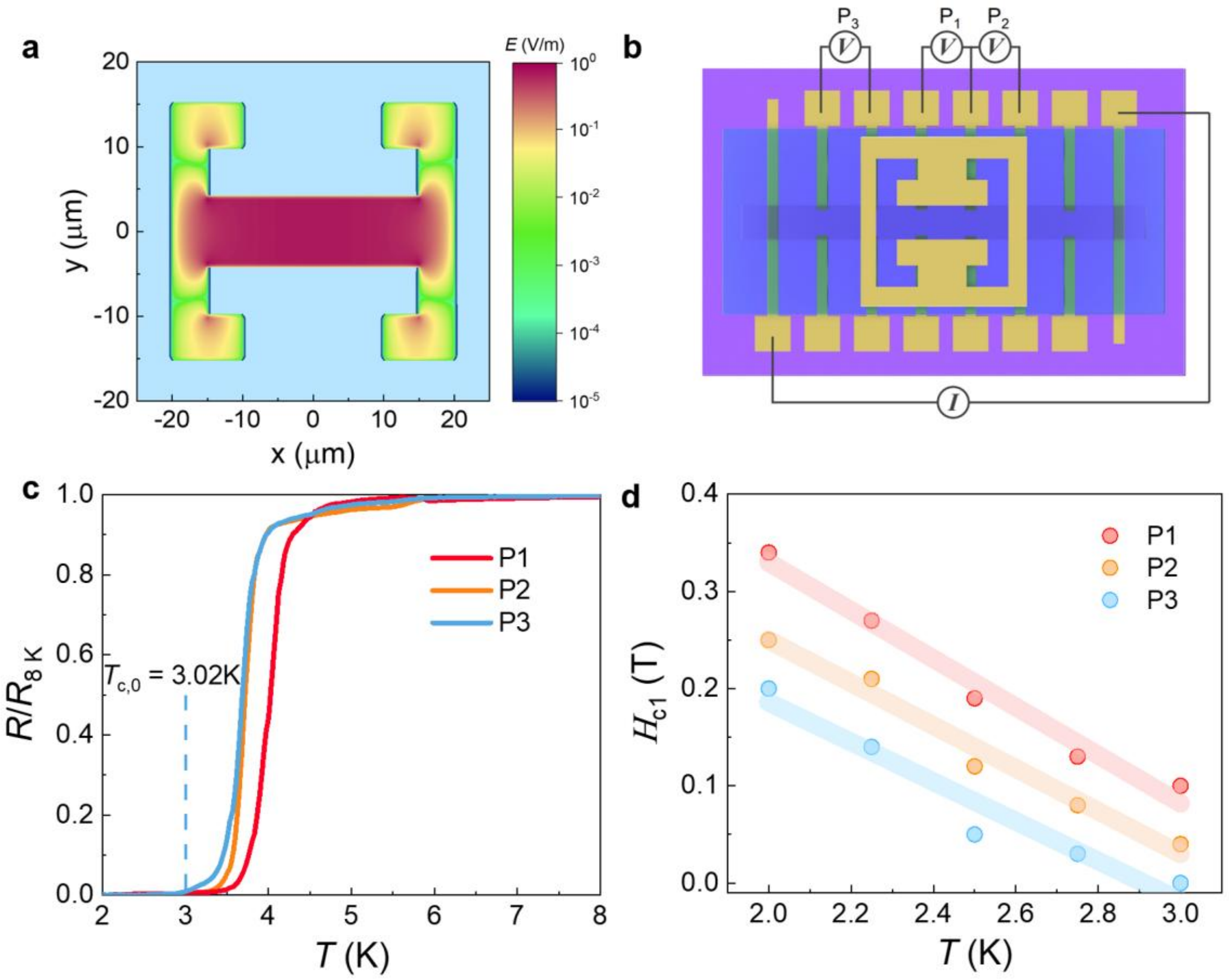

**Figure 2 | Dependence of superconducting enhancement on the local cavity electromagnetic field.** **a**, Simulated electric-field distribution of the CSRR. The field is approximately uniform (~0.8 V/m) within the cavity gap and decays rapidly outside, vanishing within ~4 μm from the gap edge. **b**, Schematic of the measurement configuration. P1, P2 and P3 refer to the center of the CSRR, the edge of the CSRR and a region outside the cavity, respectively. **c**, *R*-*T* curves measured at P1, P2 and P3. The corresponding $T_{c,0}$ values are 3.41, 3.29 and 3.02 K, respectively. **d**, Temperature dependence of the lower critical field ($H_{c1}$) measured at P1, P2 and P3.

## Figure 3

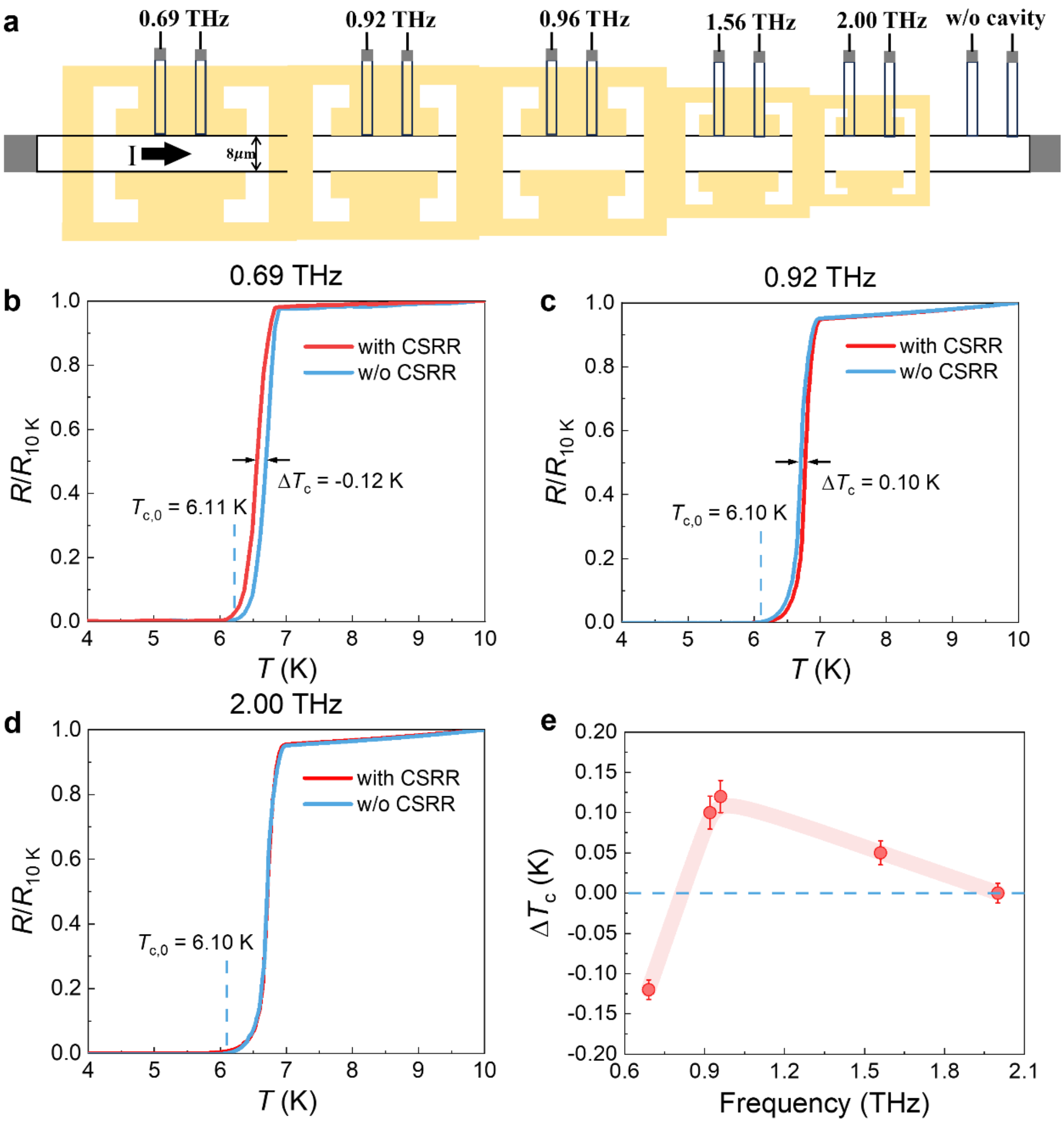

**Figure 3 | Dependence of superconducting enhancement on the frequency of CSRR. a,** Schematic of the frequency-dependent superconductivity measurements. **b-d**, *R*-*T* curves of ten-layer $NbSe_2$ devices embedded with CSRR of different resonance frequencies: **b**, 0.69 THz, **c**, 0.92 THz and **d**, 2.00 THz. Relative to the reference regions without CSRRs, the $T_{c,0}$ is suppressed by 0.12 K at 0.69 THz, enhanced by 0.10 K at 0.92 THz, and remains essentially unchanged at 2.00 THz. **e**, Summary of the change in $T_{c,0}$ as a function of cavity resonance frequency. Error bars represent the standard deviation obtained from measurements on multiple devices.

# Figure 4

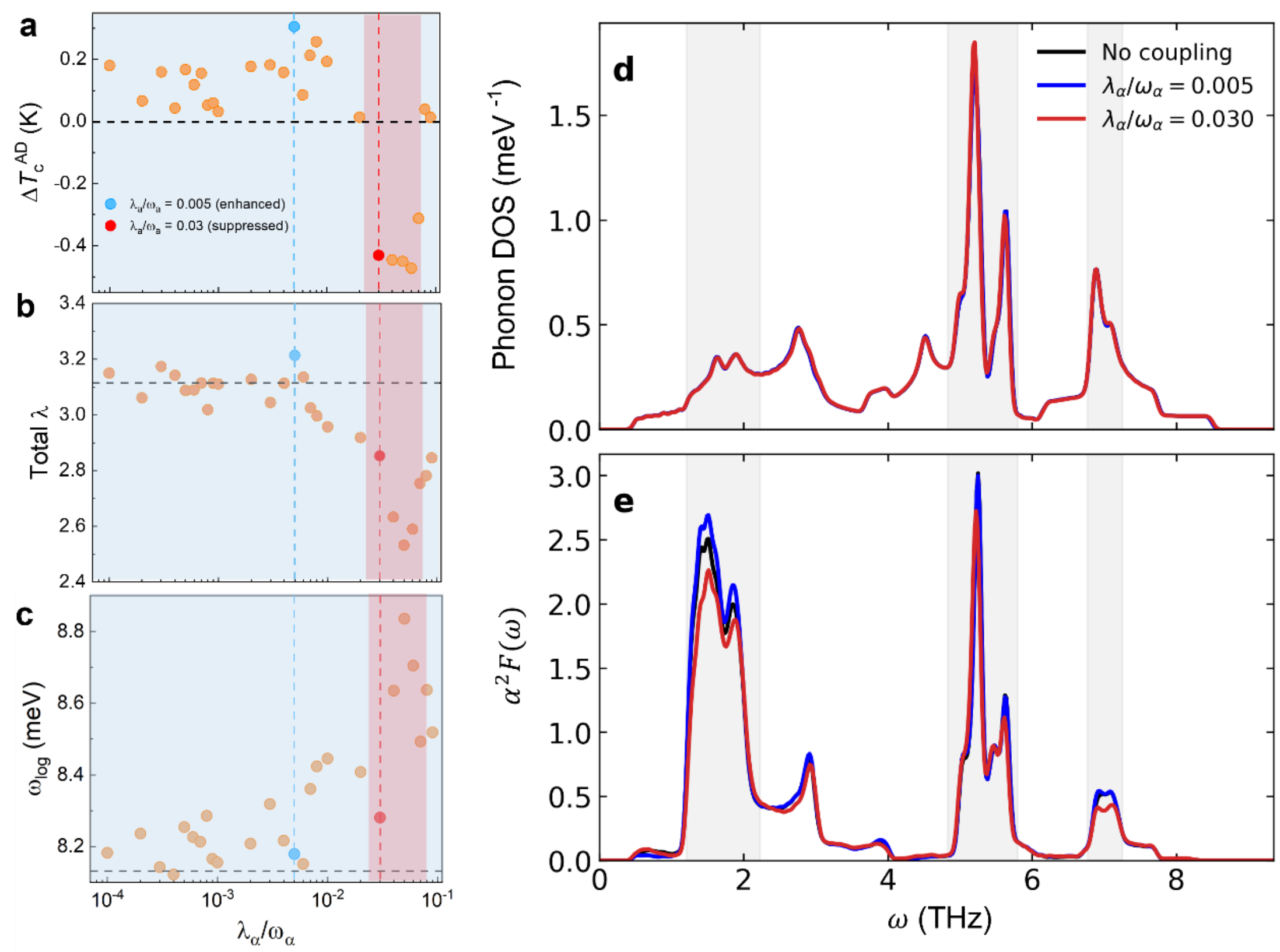

**Figure 4 | Cavity-modified electron-phonon coupling and Allen-Dynes $T_c$ in monolayer 2H-NbSe$_2$. a**, Change in the Allen-Dynes superconducting transition temperature, $\Delta T_c^{AD} = \Delta T_c^{AD}\left(\frac{\lambda_\alpha}{\omega_\alpha}\right) - \Delta T_c^{AD}(0)$, as a function of the effective photon coupling strength $\frac{\lambda_\alpha}{\omega_\alpha}$, computed for monolayer 2H-NbSe$_2$ in the non-CDW phase using $T_c^{AD} = \frac{\omega_{log}}{1.2}\exp[\frac{-1.04(1+\lambda)}{\lambda-\mu^*(1+0.62\lambda)}]$ with $\mu^* = 0.15$. Blue and red points mark the two representative coupling strengths ($\frac{\lambda_\alpha}{\omega_\alpha} = 0.005$ and $0.03$) shown in **d** and **e**, corresponding to enhanced and suppressed $T_c$, respectively; blue and red shading indicate coupling-strength regions where $\Delta T_c^{AD}$ is positive and negative, respectively. **b**, Total electron-phonon coupling strength, $\lambda = 2\int \frac{\alpha^2F(\omega)}{\omega}d\omega$ , as a function of $\frac{\lambda_\alpha}{\omega_\alpha}$. The dashed line marks the uncoupled ($\frac{\lambda_\alpha}{\omega_\alpha} = 0$) reference value. **c**, Logarithmic average phonon frequency, $\omega_{log} = \exp[\frac{2}{\lambda}\int \frac{\alpha^2F(\omega)}{\omega}\log(\omega)\, d\omega\,]$, as a function of $\frac{\lambda_\alpha}{\omega_\alpha}$, with the dashed line marking the uncoupled reference value. $\lambda$ decreases while $\omega_{log}$ increases over the same coupling range in which $\Delta T_c^{AD}$ changes sign, reflecting a competition between the two quantities in the Allen-Dynes formula. **d**, Phonon density of states (DOS) and **e**, Eliashberg spectral function $\alpha^2F(\omega)$ for the uncoupled case (black) and for the two representative electron-photon coupling strengths highlighted in **a** (blue, $\frac{\lambda_\alpha}{\omega_\alpha} = 0.005$; red, $\frac{\lambda_\alpha}{\omega_\alpha} = 0.03$).

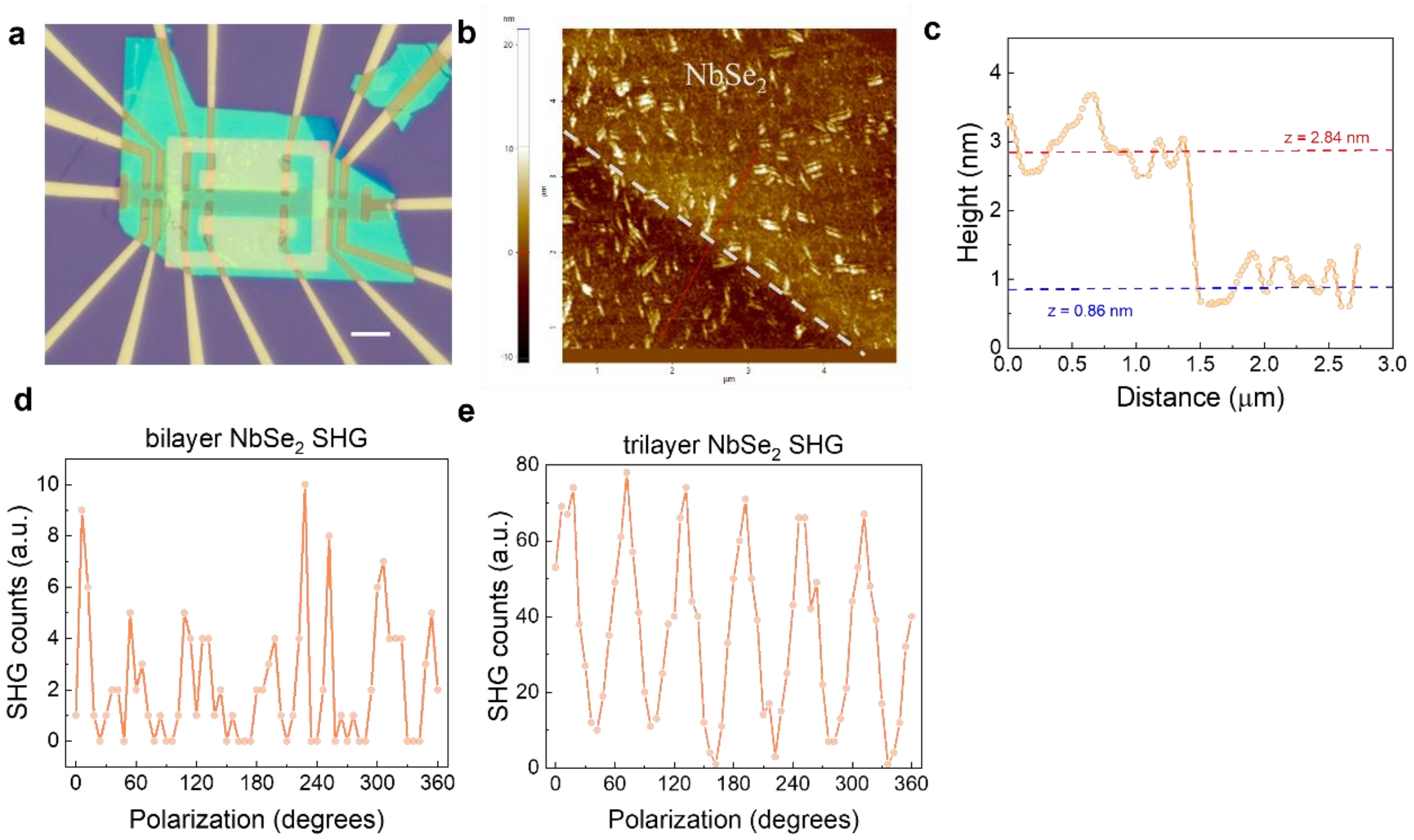

**Extended Figure 1. Sample characterization. a**, Optical image of a CSRR-embedded $NbSe_2$ device. The scale bar represents 10 μm. **b**, Atomic force microscopy (AFM) image of hBN-encapsulated $NbSe_2$. White dashed line represents edged of $NbSe_2$. **c**, Height profile of AFM scan along the red line in **b**. Thickness of the $NbSe_2$ sheet is 1.98 nm. **d**, Second-harmonic generation (SHG) signal of the sample and **e**, SHG signal of a trilayer $NbSe_2$ sample. SHG signal is reduced by an order of magnitude in bilayer $NbSe_2$ due to its restoration of inversion symmetry. Thickness of our sample is determined by combining the AFM height profile with SHG results.

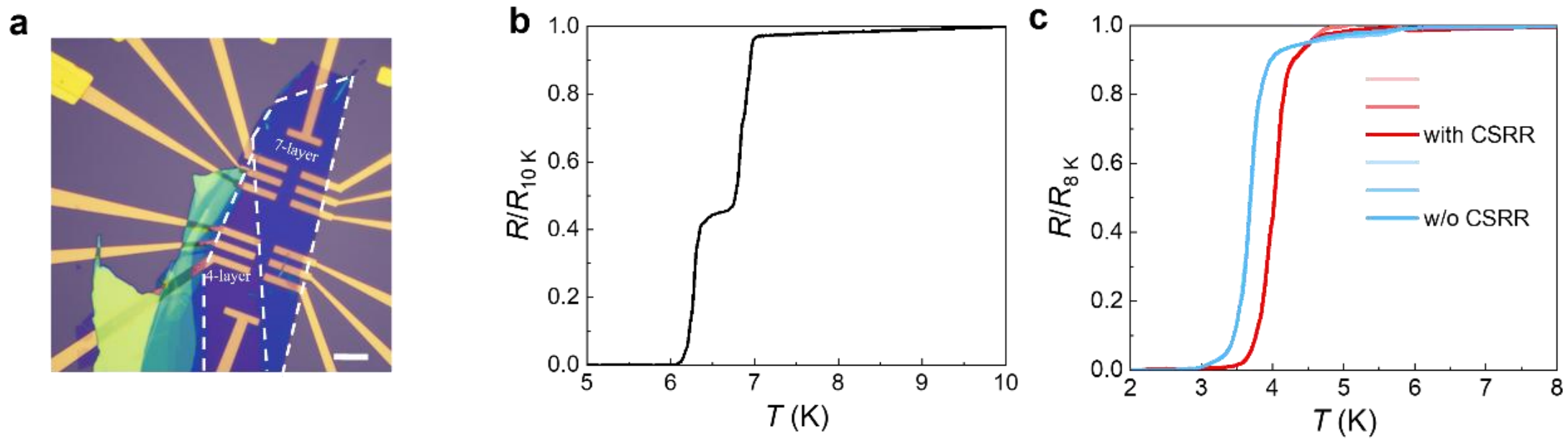

**Extended Figure 2. Excluding sample inhomogeneity as the origin of superconducting enhancement. a**, Optical image of a CSRR-embedded $NbSe_2$ device containing regions with different thicknesses (4 and 7 layers). The scale bar represents 10 μm. **b**, *R*-*T* curve measured on the device shown in **a**, exhibiting a ladder-like superconducting transition due to the coexistence of regions with different layer numbers and $T_c$. **c**, *R*-*T* curves measured with different electrode pairs in regions embedded with and without the CSRR in device shown in Figure 1.

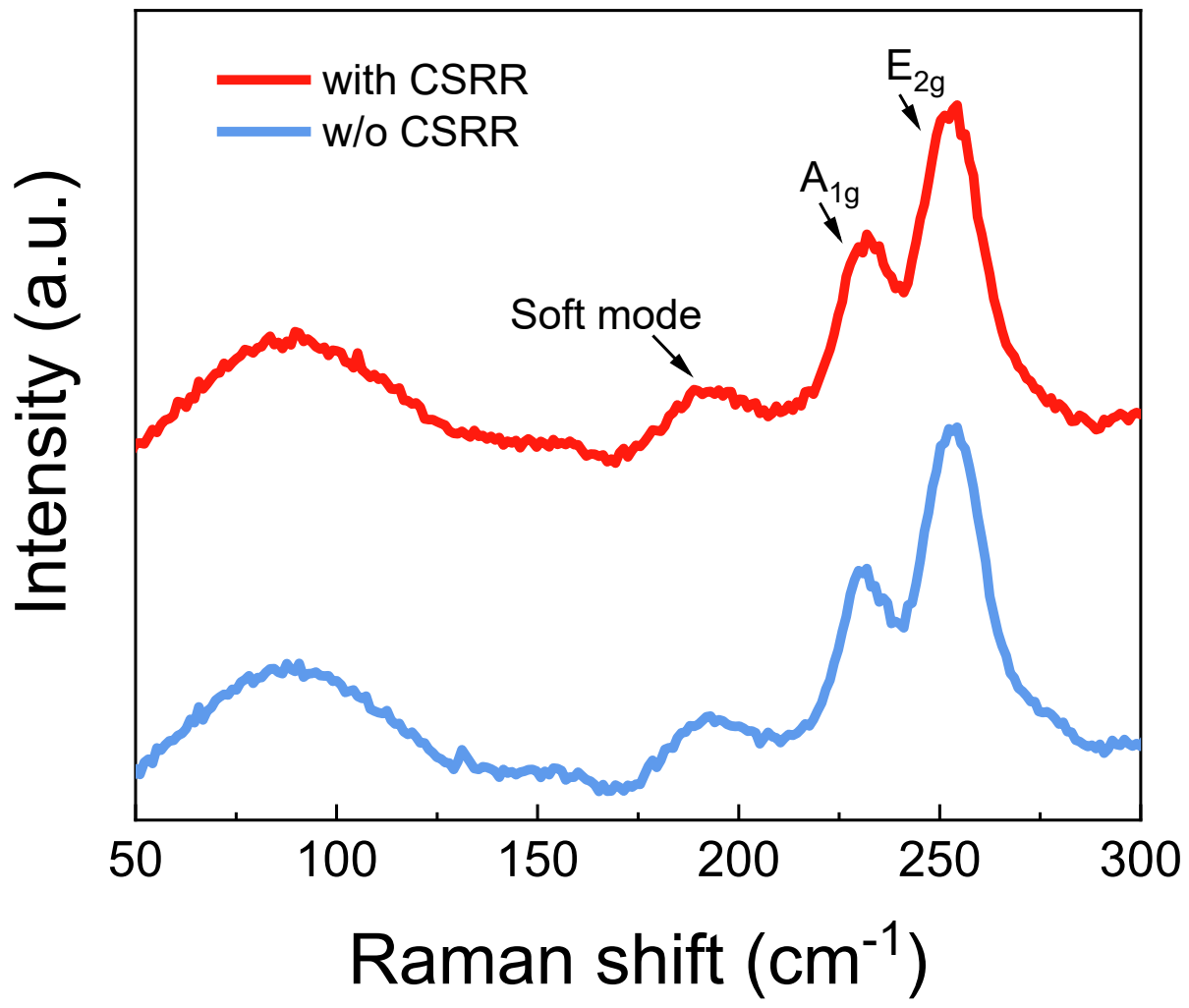

**Extended Figure 3. Raman spectroscopy**. Raman spectra measured in region with and without CSRR.

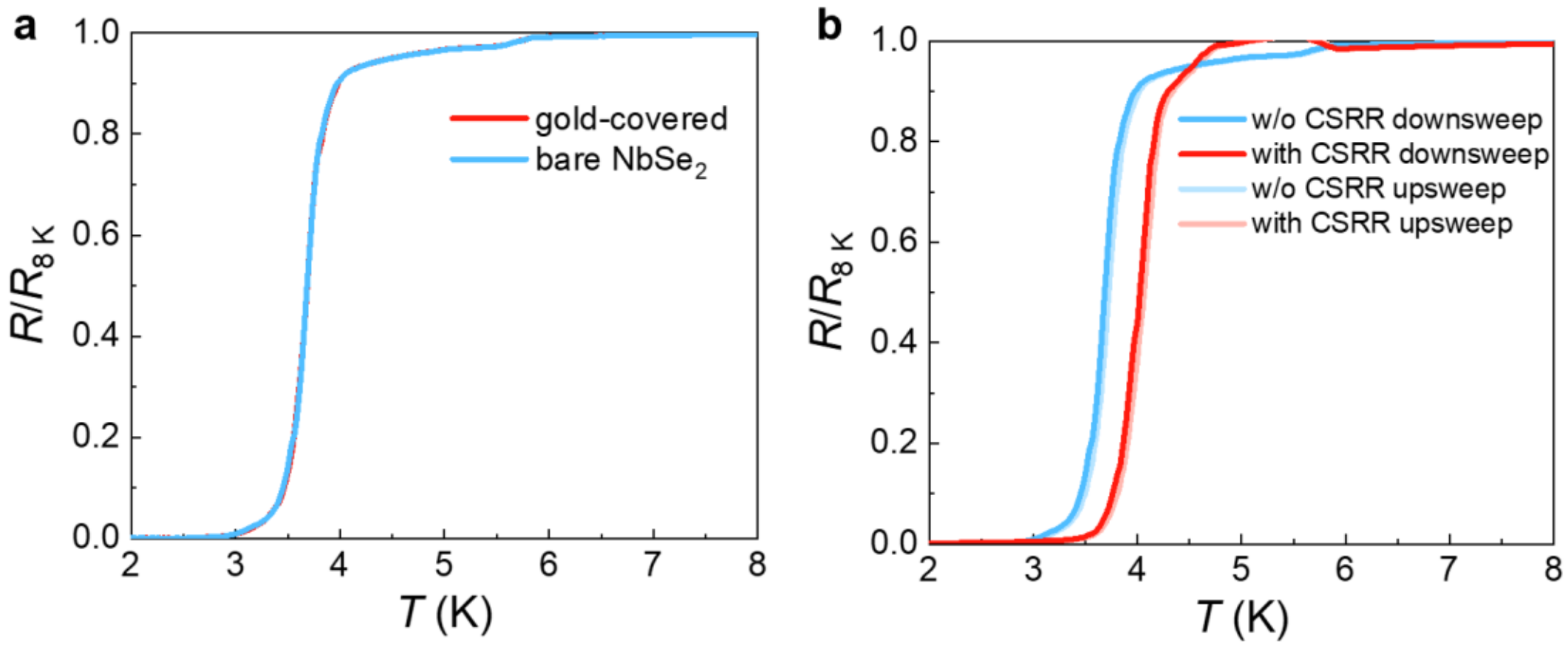

**Extended Figure 4. Ruling out contribution from gold proximity effect and thermalization.** **a**, *R*–*T* curves measured at locations covered by gold or not. Two curves are identical. **b**, Downward (dark) and upward (light) temperature sweeps in regions embedded with and without the CSRR. The downward and upward curves are almost identical with a mismatch of less than 0.02 K.

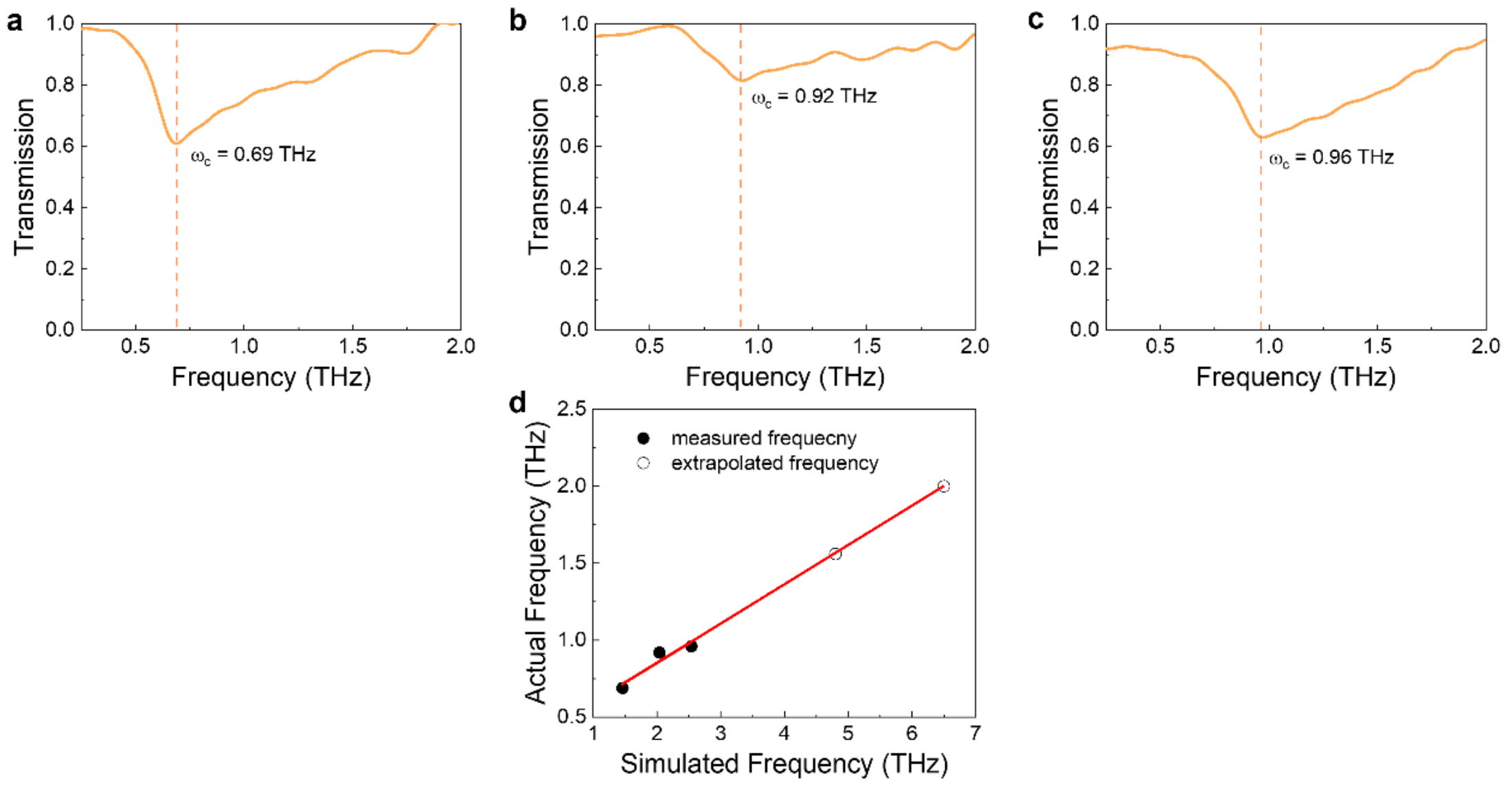

**Extended Figure 5. THz-TDS transmission spectrum of different CSRRs.** THz-TDS transmission spectrum of CSRR reveals resonant frequency of **a**, 0.69 THz, **b**, 0.92 THz and **c**, 0.96 THz. **d**, Data fitting and extrapolation of resonant frequencies of the two high frequency ones.

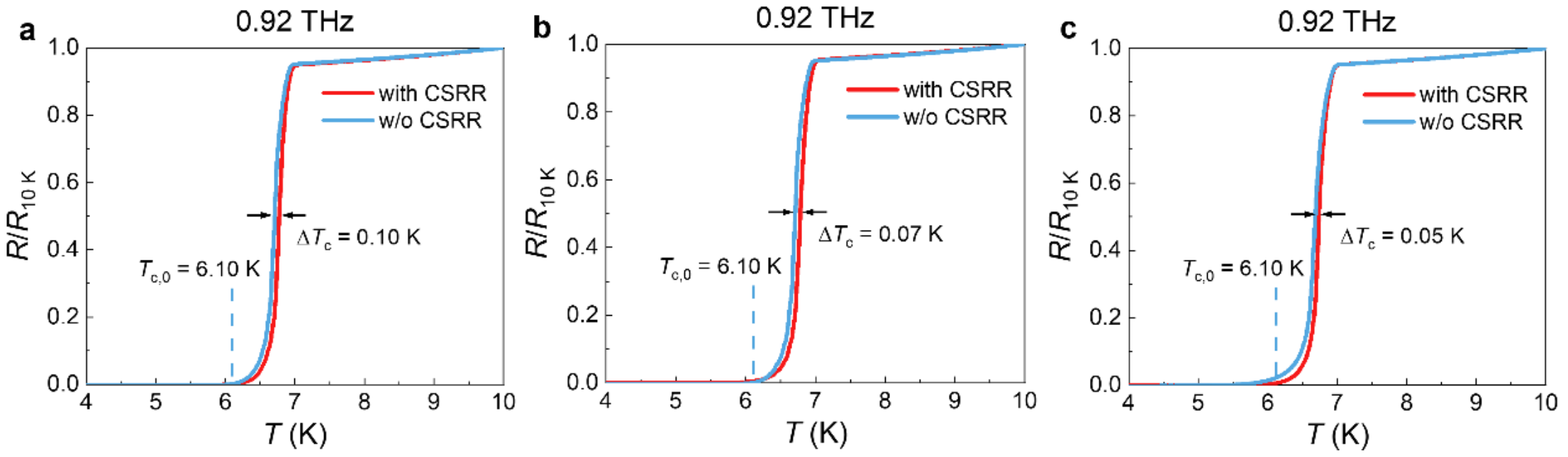

**Extended Figure 6. Reproducibility of cavity-enhanced superconductivity in $NbSe_2$ devices. a, b**, *R*-*T* curves measured at two different locations of the same $NbSe_2$ device. **c**, *R*-*T* curves measured on a second $NbSe_2$ device. A similar enhancement of the superconducting transition temperature is observed.